# Face mask perception during the COVID-19 pandemic: an observational study of Russian online social network VKontakte[1]

Alexander G. Chkhartishvili*. Dmitry A. Gubanov**. Ivan V. Kozitsin***

*V.A. Trapeznikov Institute of Control Sciences Russian Academy of Sciences, Moscow, Russia, (e-mail: sandro_ch@mail.ru)

** V.A. Trapeznikov Institute of Control Sciences Russian Academy of Sciences, Moscow, Russia, (e-mail: dmitry.a.g@gmail.com)

*** V.A. Trapeznikov Institute of Control Sciences Russian Academy of Sciences, Moscow, Russia, Moscow Institute of Physics and Technology, Dolgoprudny, Russia (e-mail: kozitsin.ivan@mail.ru)

**Abstract**: This cross-sectional study characterizes users' attitudes towards the face mask requirements introduced by the Russian government as a response to the COVID-19 pandemic. We study how they relate to other users' characteristics such as age, gender, and political attitudes. Our results indicate that men and elder individuals—demographic groups that are most vulnerable to COVID-19—underestimate the benefits of wearing face masks. We also discovered that users in opposition to the Russian government highly approve of this anti-COVID-19 measure—an oppositionist will approve of the face mask requirements with the probability of 0.95. For those who support the Russian government, the odds of approval are merely 0.45.

*Keywords*: COVID-19, online social networks, face masks wearing, risk perception, political attitudes

## 1. INTRODUCTION

The spread of COVID-19 has forced governments to impose unprecedented restrictive measures on their people. Lockdowns have been introduced in many countries, and most governments have decided to require citizens to wear masks. This measure has been proven to have a positive impact on the spread of COVID-19 (Lyu and Wehby, 2020). However, requirements alone are often not enough to effectively combat a pandemic. An important factor in this is how closely citizens comply with the restrictions imposed (Haischer *et al.*, 2020; Barrios *et al.*, 2021). From this perspective, governments essentially must understand what citizens think about the restrictions.

Online social networks (OSNs) are places where individuals can express their opinions on certain issues. The form of expression varies: individuals may write posts or comment on them. These textual messages can indicate their views. Furthermore, users may *like* or *repost* other users' posts and comments. These types of actions can be considered (indirect) manifestations of users' views. During the pandemic, a huge amount of users' attention was devoted to different pandemic-related topics such as wearing masks, vaccination, and even the issue of whether the pandemic itself exists or has been fabricated (DeVerna *et al.*, 2021; Lopreite *et al.*, 2021).

In this cross-sectional study, we analyze how users of OSN VKontakte (VK) react to posts related to COVID-19. Our main goal is to characterize users' attitudes towards the face mask requirements introduced by the Russian government. These requirements were released in the Russian Federation at different times depending on the region. In most Russian regions, it was necessary to wear protective masks in public places by 21 May, 2020.

We also investigate how users' attitudes toward the face mask requirements relate to their political preferences (we use information sources users are subscribed to as a proxy of their political attitudes) and their demographic characteristics (age and gender).

## 2. LITERATURE

Face mask–wearing has been proven to have a positive effect on limiting the spread of COVID-19 (Lyu and Wehby, 2020). A remarkable number of studies have aimed to understand how citizens perceive the face mask requirements (as well as other measures such as lockdowns or social distancing) and whether they comply with them (Haischer *et al.*, 2020; Howard, 2020, 2021). Haischer *et al.* (2020) reported that women, elder individuals, and citizens of urban areas are more likely to wear masks. In turn, Howard (2021) suggested that gender has no significant effect on whether an individual wears a face mask, but it affects how they perceive it.

A group of studies has documented that political affiliation affects individuals' risk perception and, in particular, how they perceive the effectiveness of the recommendations of the government's leaders to stay at home (Barrios and Hochberg, 2020; Grossman *et al.*, 2020). They argued that US counties that had higher rates of people voting for Donald Trump are associated with lower levels of COVID-19 risk perception. Further, state government leaders' recommendations were more effective in Democratic-leaning counties. These results are quite nontrivial since, in the face of danger, individuals must be as objective and rational as possible. As such, it is essential to understand the profound nature of this effect. For example, Barrios and Hochberg (2020) argue that political preferences largely shape the information space around individuals (due to individuals' tendency to receive information from sources congruent with their own views) and thus potentially limit their access to objective information (since information broadcasted by politically biased media may be disturbed). In OSNs, this regularity may be exacerbated due to personalization systems that can increase the limiting effect and foster the formation of echo chambers (Perra and Rocha, 2019; Kozitsin and Chkhartishvili, 2020).

## 3. METHODS

### 3.1 Data collection

We gathered data from VK, which operates rather similarly to Facebook. VK accounts can be naturally divided into two sets: information producers and information consumers. The former encompasses public pages, groups, bloggers, and event pages. Note that representatives of news outlets usually have the status of public pages. Information consumers are ordinary users that essentially have fewer followers (users with more than 1000 followers become the official status of bloggers). We focus on posts published by public pages and how ordinary VK users react to them.

We have chosen two public pages—Meduza and RT—that are representative of the corresponding news outlets.[2] These outlets are considered to hold opposing political sides: RT is traditionally associated with the Russian official government, while Meduza is considered to have much less loyalty to the Kremlin.[3] Recently, Meduza has been given the status of "foreign agent."[4] Because these two public pages have such contending political leanings, the posts published by them should be of different political sides, and the same tendency should be found in the political views of their followers: Meduza followers should be more oppositional whereas RT followers should be more conservative. It provides an opportunity to analyze individuals with potentially different political attitudes.

We gathered all posts published by these pages within a predefined period (1 April–1 June, 2020) that contained at least one word from a predefined set of words from the topic of COVID-19 (accounting for different word forms): "ковид," "коронавирус," "covid," "coronavirus," "карантин," "удаленк," "самоизоляц," "пандеми," and "эпидеми." Next, we downloaded all comments related to these posts, as well as likes left by users on these posts and comments. In this conference paper, we concentrate only on a small subset of the information cascades induced by these posts. More precisely, we included individuals who took part in these information cascades by commenting the posts with words that included at least one word from the following: "маск," "масочн," and "намордник" (these words relate to the topic of face mask requirements). By doing so, we collected a sample $I_c$ of 1559 users. We denote the set of corresponding comments as $C$.

We classified each comment from the set $C$ according to the attitude it conveyed toward the face mask requirements. We highlighted the following cases: (1) negative attitude; (2) neutral attitude or attitude is not provided; and (3) positive attitude. Note that a negative attitude does not necessarily mean that an individual does not wear a face mask; it describes only how they perceive face mask–wearing. Using this classification, for each user $i$ from the sample $I_c$ we identified $i$'s attitude towards the face mask requirements manually by labeling them with the variable $a_i^c \in \{-1,0,1\}$. This variable is referred to as the *comment*-attitude (meaning that the attitude stems from the user's comments).

We further classified individuals who had liked at least one comment from the set $C$ with a negative or positive attitude. We denoted the set of such users with $I_l$ ($|I_l| = 847$). Each user $j \in I_l$ was associated with the variable $a_j^l \in \{-1,1\}$ that stems from the attitudes of comments liked by $j$ (the *like*-attitude).

We focused on users from the set $I = I_c \cup I_l$. The sets $I_c$ and $I_l$ had 113 intersections—users who both made at least one comment on the topic of face mask requirements (under the posts on the topic of COVID-19) and liked at least one such comment with a positive or negative attitude. We carefully processed all these instances. We found 10 users who had liked a comment with an orientation opposite to what they had written themselves. Excepting these individuals, each user $i$ from $I$ was assigned the variable $a_i \in \{-1,0,1\}$ representing their *overall* attitude towards the face mask requirements. The detailed scheme of how we defined the variable $a_i$ can be found in Table 1. Note that the situation where $a_i = 0$ may correspond to qualitatively different cases: it may relate to a user with a neutral attitude, or it may stand for a user that has a positive or negative attitude, but this cannot be inferred from the data.

**Table 1. Users' attitudes towards the face mask requirements manifested by comments ($a_i^c$) and likes ($a_i^l$)**

| Comment-attitude $a_i^c$ | Like-attitude $a_i^l$ | Overall attitude $a_i$ | Number of users |
|---|---|---|---|
| -1 | None | -1 | 106 |
| -1 | -1 | -1 | 10 |
| -1 | 1 | Not defined | 7 |

---

[2] https://meduza.io/en, https://www.rt.com/
[3] https://archives.cjr.org/feature/what_is_russia_today.php, https://www.rt.com/op-ed/377715-rt-gazillionaire-media-funding/
[4] https://meduza.io/en/feature/2021/04/30/save-meduza

|   |   |   |   |
|---|---|---|---|
| 0 | None | 0 | 1098 |
| 0 | -1 | -1 | 15 |
| 0 | 1 | 1 | 41 |
| 1 | None | 1 | 242 |
| 1 | -1 | Not defined | 3 |
| 1 | 1 | 1 | 37 |
| None | -1 | -1 | 178 |
| None | 1 | 1 | 556 |
| Total number of users | | | 2293 |

For individuals from $I$, we obtained their age and gender (0: female; 1: male), as well as information on their connections to other accounts on VK. We considered two types of connections: (1) friendship (reciprocated) ties with users and (2) following (unreciprocated) ties directed on bloggers and public pages. Among the sample (consisting of 2293 individuals) we found only six friendship connections: four connected pairs and one open triad. It ensures that the observations are independent (since in social networks, neighboring vertices' characteristics are usually correlated).

*3.2 Users' political attitudes*

We estimated the political attitudes of users from $I$ by determining how oppositional users are with respect to the Russian government. Each user $i$ was associated with the continuous variable $x_i \in [0,1]$ denoting the extent to which user $i$ is inclined to support the opposition movement. $x_i = 1$ indicates that $i$ is a strong oppositionist whereas $x_j = 0$ means that $j$ is loyal to the government. For simplicity, we define $j$ as a conservative user. We calculated $x_i$ using a slightly modified version of the algorithm presented by Kozitsin *et al.* (2020). Simply, $x_i$ is a projection from the set of information sources (bloggers and public pages) that user $i$ is subscribed to. It reflects the fact that users tend to follow information sources that have a political bias congruent with their own views (Frey, 1986). As such, $x_i = 1$ means that $i$ follows highly oppositional information sources whereas $x_j = 0$ indicates that $j$ is subscribed to conservative public pages and bloggers.

*3.3 Additional filters*

In our analysis, we impose additional constraints. We considered only users with a well-defined overall attitude towards the face mask requirements (i.e., we did not consider those with $a_i = 0$). Furthermore, we considered only accounts that had more than 10 subscriptions to public pages and bloggers. This filter was proposed by Kozitsin *et al.* (2020), who discovered that estimations of the political attitudes of such users were of the highest accuracy. Following these adjustments, we were left with 582 observations.

## 4. RESULTS

Our analysis revealed that the sample of interest was comprised of slightly more male users (see Fig. 1). Only 254 users noted their age (about half) and nine of them had anomalously high values of age (greater than 90 years), which were likely fabricated. Most users were estimated to have moderate political attitudes; the relatively small peaks at 0 and 1 correspond to radical conservatives and oppositionists, respectively. We also see that the number of people who supported the face mask requirements was more than twice that of those who had a negative attitude toward wearing masks.

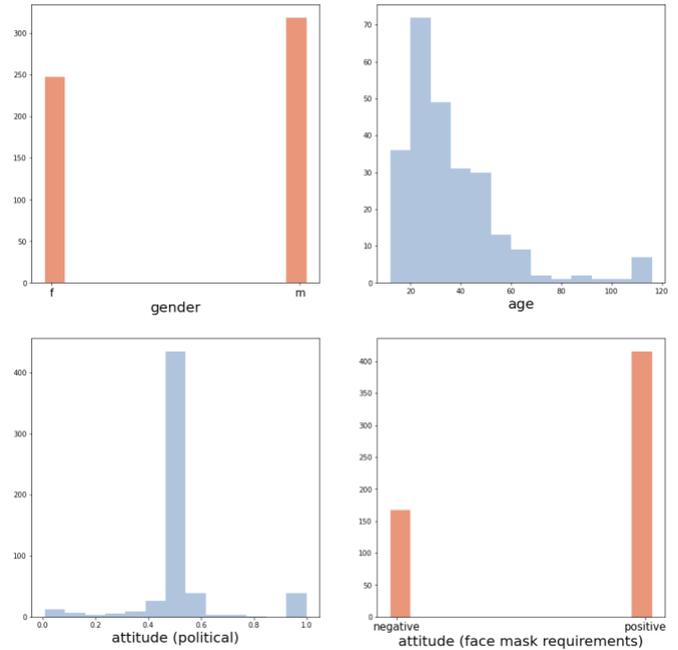

Figure 1. Histograms displaying users' characteristics.

Next, we studied how individuals' attitudes toward the face mask requirements were connected to their other characteristics. We specifically questioned how these attitudes vary along political lines. Similarly to Barrios and Hochberg (2020) and Grossman *et al.* (2020), we report that attitudes toward the face mask requirements varied strongly across political lines (see Fig. 2). The main trend observed is that the more oppositional a user is, the higher the odds are that they approve of the face mask requirements.

We performed regression analysis in which we used $a_i$ as the dependent variable and gender, age, number of friends, and $x_i$ as independent variables. In this analysis, we ignored users aged over 90. After fitting the logit model on the remaining instances (see Table 1), we found that the tendency of oppositional users to have a higher rate of approval for wearing face masks holds even after controlling for other characteristics. We also found that an increase in age or being male leads to lower odds of face mask approval, other things being equal. The number of friends had no significant effect on the dependent variable.

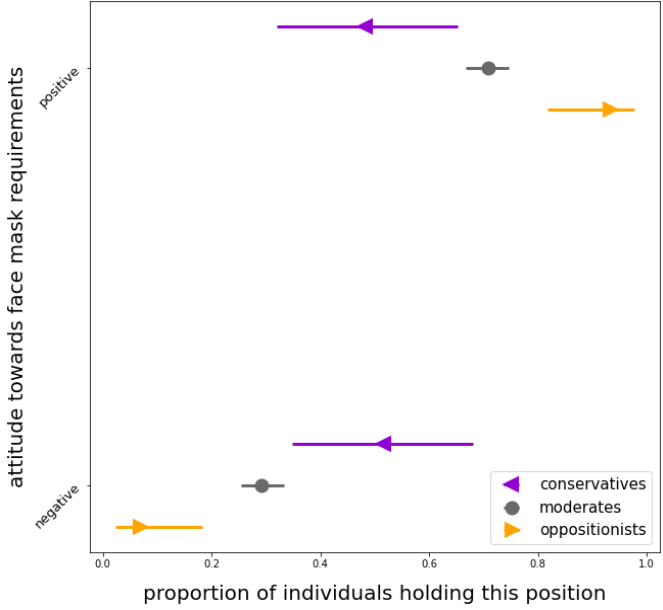

Figure 2. The proportions of individuals having different attitudes towards the face mask requirements. The groups themselves are constructed as follows: conservatives are users with $x_i < 0.33$, oppositionists are characterized by $x_i > 0.66$, moderates are those in between.

Table 2. Logistic regression summary

| Coefficient | Estimate | Standard error | Pr(>\|z\|) |
|---|---|---|---|
| Intercept | 0.309 | 0.791 | 0.696 |
| Age | -0.022 | 0.01 | 0.036* |
| Gender | -0.751 | 0.325 | 0.021* |
| Number of friends | 0.011 | 0.008 | 0.148 |
| Political attitude $x_i$ | 3.431 | 1.237 | 0.006** |

Signif. codes: 0 '***' 0.001 '**' 0.01 '*' 0.05 '.' 0.1 ' ' 1

The effect of demography characteristics confirms the irrational nature of individuals' perceptions since men and elder individuals are the demographic groups that are considered to be most vulnerable to COVID-19. Note that our results do not contradict those of Haischer *et al.* (2020), who argued that elder individuals are more prone to wear face masks: unlike that study, we analyze users' attitudes, not their actions. The way political attitudes (and more precisely, information embedding) affect attitudes toward wearing face masks is quite surprising since the oppositionists should, a priori, negatively perceive all measures taken by the government. Conversely, those who highly approve of the government should also approve orders released by the government. We leave this puzzle for future research.

## 5. PURE-TYPES MODEL

In this section, we develop a probabilistic model that fixes the relationship between pure types (conservative/oppositionist; reject mask-wearing/approve of mask-wearing) and derive the method of estimating its parameters. We further illustrate how this method works on the data at hand.

Recall that the quantity $x_i$ characterizes users on a continuous scale. However, decisions made by individuals imply a finite number of alternatives. The simplest example is an election with two parties (Proncheva, 2020). From this perspective, the variable $x_i$ can be conceptualized as the probability that, in a particular situation, user $i$ will act as an oppositionist (conversely, $1 - x_i$ stands for the chance of acting as a conservative). Next, let us assume that an oppositionist agrees with the face mask requirements with a fixed probability $p$ and a conservative person with probability $q$. Here we do not explain the nature of these probabilities: it may be caused by some sort of information distortion or something else. We suppose that $p$ and $q$ are determined solely by the individual's type (conservative or oppositional) and are not sensitive to external effects such as, say, peer influence processes (but this assumption does not hold for the probability $x_i$ of being oppositional [Gubanov and Chkhartishvili, 2015; Kozitsin, 2020]). The data at hand provide an opportunity to obtain quantities $p$ and $q$ using maximum likelihood estimation. The probability of complying is defined by $x_i p + (1 - x_i)q$. In turn, the probability of rejecting is given by $x_i(1 - p) + (1 - x_i)(1 - q)$. As such, we can write the likelihood function:

$$L(p,q) = \prod_{i \in I_+} [x_i p + (1 - x_i)q] * \prod_{i \in I_-} [x_i(1 - p) + (1 - x_i)(1 - q)]$$

where $I_+$ are individuals from the filtered sample with $a_i = 1$ and $I_-$ are those for whom $a_i = -1$. To estimate the values of $p$ and $q$, one should solve the following optimization problem:

$$\min_{\delta \leq p, q \leq 1-\delta} -\ln L(p, q),$$

where $\delta$ is a small positive constant (we set it as $10^{-5}$) that aims to avoid situations when multipliers under the sign of the logarithm are null. We solve this optimization problem using the function *minimize* introduced in SciPy package with the Sequential Least Squares Programming (SLSQP) method.[5] This approach requires providing an initial guess point. Our analysis reveals that for almost all initial points (excepting those of (0,0) and (1,1)) the result is largely the same: $\hat{p} = 0.95$, $\hat{q} = 0.45$. The value of $\delta$ does not have a significant impact on the outcome of the optimization algorithm.

---

[5] https://docs.scipy.org/doc/scipy/reference/optimize.minimize-slsqp.html#optimize-minimize-slsqp

Estimated values of $p$ and $q$ indicate that, with a probability of 0.95, oppositionists will approve the facemask requirements whereas conservatives will do so with a probability of 0.45. That is, oppositionists approve face mask–wearing at an extremely high rate while conservatives act approximately at random.

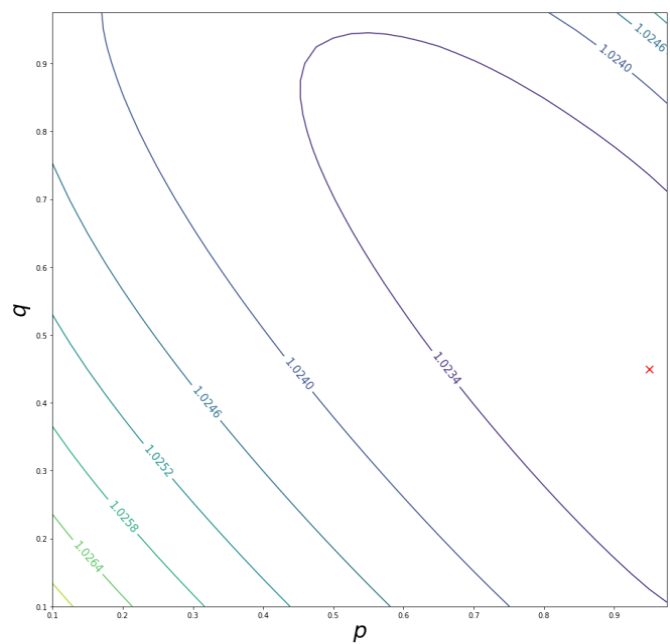

Figure 3. Contour lines of the function $(-\ln L(p,q))^{1/256}$. The red "x" marks the minima obtained.

## 6. DATA AVAILABILITY

All data used in this study can be obtained upon reasonable request.

## 7. FUNDING

This work was partially supported by the Russian Foundation for Basic Research, project no. 20-04-60296.

## 8. CONCLUSION

This cross-sectional study characterizes users' attitudes towards the face mask requirements introduced by the Russian government as a response to the COVID-19 pandemic. We study how they relate to other users' characteristics such as age, gender, and political attitudes.

Our research can be useful in elaborating the government's anti-COVID-19 policies as it demonstrates that different demographic and ideological groups may perceive the same measures in different ways.

Possible directions for future research include analysis of the coevolution of political and pandemic-related attitudes, which may provide more understanding of how these kinds of attitudes are connected to each other and what the role of demographic characteristics is in this process (Peshkovskaya, Babkina and Myagkov, 2019). Another interesting avenue for research could be the analysis of the data from the perspective of multidimensional opinion formation models (Parsegov et al., 2017; Gubanov and Petrov, 2019; Howard, 2020; Petrov and Proncheva, 2020).